 \documentclass[final,5p,times,twocolumn,number,sort&compress]{elsarticle}

\usepackage{dcolumn}
\usepackage{bm}
\usepackage{amsmath}
\usepackage{amssymb}
\usepackage{lipsum}
\usepackage{doi}
\usepackage{url}
\usepackage{booktabs}

\journal{Physics Letters A}

\begin{document}

\begin{frontmatter}



\title{Minkowski and Abraham Momenta Revisited from the Perspective of Photon Dynamics}

\author[first]{Lili Yang\corref{cor1}}
\ead{yanglli7@mail.sysu.edu.cn}
\author[first]{Pengming Zhang}
\author[first]{Longlong Feng}
\cortext[cor1]{Corresponding author}

\affiliation[first]{
            organization={School of Physics and Astronomy, Sun Yat-Sen University},
            addressline={}, 
            city={Zhuhai},
            postcode={519082}, 
            state={},
            country={China}}

\begin{abstract}
The distinction between Minkowski and Abraham momenta has traditionally been discussed in the context of electron dynamics. In this work, we revisit the problem from the perspective of photon dynamics in a medium, and show that, within a field-theoretic formulation, their relation is naturally encoded through covariant and ordinary derivatives, together with a medium-induced gauge field. In this quantized framework, the microscopic origin of the optical Hall effect is identified as a spin–Zeeman–type coupling that emerges from the Dirac-like equation for photons.
\end{abstract}




\end{frontmatter}

\section{Introduction}

The momentum density of an electromagnetic field in a material medium is commonly described by either the Minkowski form $\mathbf{S}^{\rm Min}=\mathbf{D}\times\mathbf{B}$ or the Abraham form $\mathbf{S}^{\rm Abr}=\mathbf{E}\times\mathbf{H}$. The corresponding total momenta are acquired by volume integration of the momentum density, namely $\mathbf{P}^{Min} =\int d^3 r \mathbf{S}^{Min}\,$, $\mathbf{P}^{Min} =\int d^3 r \mathbf{S}^{Min}$. These two expressions have been the subject of a long-standing debate regarding their physical interpretation and applicability\cite{BRADSHAW2010650,Balazs,Padgett,Barnett3,Ramos:2013som,PhysRevA.102.063510,Ortega,RevModPhys.79.1197,10.1098/rsta.2009.0207,MANSURIPUR20101997,PhysRevA.95.063850}, with both having received experimental support \cite{Campbell,She}. Barnett and coworkers \cite{Barnett1,Barnett2} argued that the Abraham and Minkowski momenta are both physically meaningful, corresponding respectively to the kinetic and canonical momenta of the electromagnetic field in a medium. In their analysis, using the point-dipole approximation for atoms in external electromagnetic fields, they show that for a collection of such dipoles forming a medium, the total momentum of the medium and the fields is conserved $\mathbf{p}_{med}^{kin} +\mathbf{p}^{Abr} = \mathbf{p}_{med}^{can} + \mathbf{p}^{Min}$\cite{Barnett2}. Thus, the Minkowski momentum can be regarded as the canonical momentum, while the Abraham momentum is the kinetic momentum. This viewpoint offers a natural resolution of the Abraham--Minkowski controversy in terms of total momentum conservation. Their conclusion, however, was obtained from the dynamics of atoms interacting with electromagnetic fields under the dipole approximation.

From the photon perspective, we show that whether the Minkowski and Abraham momenta should be identified as canonical or kinetic momenta is representation dependent. This identification hinges on whether one quantizes the electromagnetic field together with the medium as a coupled whole, or quantizes the electromagnetic field itself. In the $\widetilde{\psi}$ representation, where the photon wave function is constructed from the electric displacement field $\mathbf D$ and the magnetic induction $\mathbf B$, the Minkowski momentum is generated by an ordinary derivative and is therefore identified as the canonical momentum under the quantum--classical correspondence, while the Abraham momentum is generated by a covariant derivative and is identified as the kinetic momentum. In the $\psi$ representation, where the photon wave function is constructed from the electric field $\mathbf E$ and the magnetic field $\mathbf H$, the roles are interchanged: the Abraham momentum is canonical, whereas the Minkowski momentum is kinetic.

\section{Representation-Dependent Forms of the Minkowski and Abraham Momenta}

The photon wave function $\widetilde{\psi}$ provides a convenient framework in which the electromagnetic-field momentum can be written in a form analogous to the derivative operators in quantum mechanics \cite{Enk1994SpinAO,Kobe1999ARS,Keller2005OnTT,BialynickiBirula1994OnTW}. It is convenient to formulate the discussion in helicity space. The transverse helicity basis vectors $\mathbf{e}_{\pm}$ are defined as the eigenvectors of the spin projection along the propagation direction, taken here to be the $z$-axis. They satisfy $(\mathbf{e}_z\cdot \mathbf{s} )\mathbf{e}_{\sigma}= \sigma \mathbf{e}_{\sigma}$, where $\sigma=\pm$ labeling the two helicities. The spin operator s is represented in the adjoint representation, i.e., $(s_i)_{jk} = -i\epsilon_{ijk}$. They satisfy the orthonormality relation $\mathbf{e}_{\sigma} \cdot \mathbf{e}^*_{\sigma'} = \delta_{\sigma\sigma'}$ and are related to the Cartesian basis by $\mathbf{e}_{\pm}=(\mathbf{e}_x\pm i\mathbf{e}_y)/\sqrt{2}$. Projecting these basis vectors onto the helicity space, the two helicity basis states can be represented as $\mathbf{e}_{+}=(1,0)^T$ and $\mathbf{e}_{-}=(0,1)^T$. Accordingly, in the helicity representation, the four-component photon wave function is defined as
\begin{equation}\label{wavefunction}
    \widetilde{\psi} =\left(\begin{array}{l}
         \widetilde{\psi}^+  \\
          \widetilde{\psi}^-
    \end{array}\right),\quad \widetilde{\psi}^{\pm} = D_{\perp} \pm i\sigma_3 B_{\perp},
\end{equation}
where $D_{\perp}$ and $B_{\perp}$ are the projections of the transverse components of $\mathbf{D}$ and $\mathbf{B}$ onto the helical basis vectors $\mathbf{e}_{\pm}$. The source-free conditions, $\boldsymbol{\nabla}\cdot\mathbf{D}=0$ and
$\boldsymbol{\nabla}\cdot\mathbf{B}=0$, constrain the longitudinal components of the fields. In particular, they can be written in terms of the transverse components as
\begin{equation}
\label{transverse}
    D_z = \frac{-1}{k_z} (\hat{k}_+D_- + \hat{k}_- D_+ ),\quad B_z = \frac{-1}{k_z} (\hat{k}_+B_- + \hat{k}_- B_+ ).
\end{equation}
Here $\hat{k}_{\pm}=(\hat{k}_x\mp \hat{k}_y)/\sqrt{2}$ are the transverse momentum operators in the chosen basis, with $\hat{k}_i=-i\partial_i$, whereas $k_z$ denotes the longitudinal component of the wave vector. Thus, even for structured light propagating predominantly along the z axis in vacuum, the local wave vector need not be parallel to the z axis; equivalently, the field may possess nonzero longitudinal components \cite{Bliokh_2015,Bliokh2013ExtraordinaryMA}.

In the leading paraxial order, the photon wave function is expanded in terms of photon and antiphoton creation and annihilation operators, analogous to the electronic case\cite{Williams2022IntroductionTQ,yang}:
\begin{equation}\label{bohanshu zhankai}
\begin{aligned}
\widetilde{\psi}(\mathbf{r}, t)
= \int d^3 k \, \frac{1}{N(k)} \sum_s
\Big[
& a_{\sigma}(\mathbf{k}) \mathbf{e}_{\sigma}(\mathbf{k})
e^{-i\omega t + i\mathbf{k}\cdot \mathbf{r}}  \\
&+ b_{\sigma}^\dagger(\mathbf{k}) \widetilde{\mathbf{e}}_{\sigma}(\mathbf{k})
e^{i\omega t - i\mathbf{k}\cdot \mathbf{r}}
\Big],
\end{aligned}
 \end{equation}
where $N(k)=\sqrt{2E_k(2\pi)^3}$ is the normalization factor and $E_k=\omega$. Here, $a_{\sigma}^{\dagger}$ and $b_{\sigma}^{\dagger}$ denote the creation operators for photons and antiphotons, respectively, while $\sigma=\pm1$ labels the photon helicity. The four-component spinors $\mathbf{e}_{\sigma}(\mathbf{k})$ and $\widetilde{\mathbf{e}}_{\sigma}(\mathbf{k})$ represent the positive- and negative-energy helicity eigenstates, respectively. Since these excitations are bosonic, the nonvanishing commutation relations are
\begin{equation}
    \begin{aligned}\label{commutation relation}
        [a_s(\bm k),a^{\dagger}_{s^{\prime}}(\bm k^{\prime})]= (2\pi)^3\delta_{ss^{\prime}}\delta (\bm k-\bm k^{\prime}) = [b_s(\bm k),b^{\dagger}_{s^{\prime}}(\bm k^{\prime}) ].
    \end{aligned}
\end{equation}
Consequently, the Minkowski momentum can be written in a derivative-operator form and, upon second quantization, takes the form with $c=\hbar=1$ \cite{yang} 
\begin{equation}\label{Min-momentum}
    \mathbf{P}^{Min}  = \int d^3r \widetilde{\psi}^{\dagger} \gamma^0 (-i\boldsymbol{\nabla}) \widetilde{\psi}=\int d^3 k \sum_{\sigma}\,\mathbf{k}\, (a^{\dagger}_{\sigma} a_{\sigma} + b^{\dagger}_{\sigma} b_{\sigma}).
\end{equation}

Here, we quantize the electromagnetic field in the medium as a unified whole within the photon wave-function formalism. We show that, in the representation based on the wave function $\widetilde{\psi}$ the Abraham momentum $\mathbf{P}^{Abr}$ takes a form analogous to a covariant derivative and is related to the Minkowski momentum through a medium-induced gauge field.  Alternatively, if the photon wave function $\psi$ is constructed from $\mathbf{E}$ and $\mathbf{H}$, in analogy with Eq.~(\ref{wavefunction}), the momentum associated with the ordinary derivative operator in Eq.~(\ref{Min-momentum}) is identified as the Abraham momentum $\mathbf{P}^{Abr}$.

\subsection{Local Minkowski and Abraham Momentum Densities in an Inhomogeneous Medium}

In the medium, the electromagnetic fields are related by the constitutive relations $\boldsymbol{\epsilon}^{-1} \mathbf{D} =\mathbf{E}$, $ \boldsymbol{\mu}^{-1} \mathbf{B} =\mathbf{H}$.
For convenience, we consider a spin-degenerate medium, for which the inverse permittivity and permeability tensors are assumed to take the same form, $\boldsymbol{\epsilon}=\boldsymbol{\mu}$. In the helicity space, the inverse permittivity tensor can be written as \cite{Feng2022}
\begin{equation}\begin{aligned}
        \boldsymbol{\mu}^{-1} = \boldsymbol{\epsilon}^{-1} =\left(\begin{array}{cc}
       \boldsymbol{\epsilon}^{-1}_{\perp}  & \mathbf{q} \\
        \mathbf{q}^{\dagger} & q_0
    \end{array}\right), \qquad  \boldsymbol{\epsilon}^{-1}_{\perp} = \left(\begin{array}{cc}
      Q_0  & Q_{+2} \\
       Q_{-2} & Q_{0}
   \end{array}\right).
\end{aligned}
\end{equation}
where the decomposition leads to two monopole moments $Q_0=\frac{1}{2}(\epsilon^{-1}_{11}+\epsilon^{-1}_{22})$ and $q_0=\epsilon^{-1}_{33}$, dipole $Q_{\pm 1}=\frac{1}{\sqrt{2}}(\epsilon^{-1}_{13}\mp i\epsilon^{-1}_{23})$ that form a dipole vector ${\bf q}=\{Q_{+1},Q_{-1}\}$, and quadrupole $Q_{ \pm 2}=\frac{1}{2}\left[\left(\epsilon^{-1}_{11}-\epsilon^{-1}_{22}\right) \mp i\epsilon^{-1}_{12}\right]$, corresponding to the scalar (spin-0), vector (spin-1), and tensor (spin-2) modes, respectively\cite{Feng2022}.

As an example, consider an impedance-matched medium with refractive index $n(\mathbf r)$, for which $\epsilon=\mu =n$. In this case, the only nonzero parameter is $Q_0=1/n $, while all other components vanish. For a homogeneous medium, the familiar relation is readily recovered. By substituting the wave-function transformation $Q_0\widetilde{\psi}=\psi$ into Eq.~(\ref{Min-momentum}), one obtains the standard Minkowski--Abraham relation, $\mathbf{P}^{Min} =n^2\mathbf{P}^{Abr}$. An equivalent derivation can be carried out in the $\psi$ representation.

In an inhomogeneous medium, the spatial variation of the constitutive parameters gives rise to additional gradient-dependent contributions to the momentum. These terms characterize the exchange of momentum and energy between the electromagnetic field and the medium and vanish in the homogeneous limit. In the $\widetilde{\psi}$ representation, the coupled photon--medium system is quantized as a whole, and the relation between the wave vector and the frequency is valid only locally, namely, $n(r)\omega =k$. To illustrate this point, we consider a plane wave in the $\widetilde{\psi}$ representation, $\widetilde{\psi}=\widetilde{\psi}_0 e^{i\mathbf{k}_{re}\cdot \mathbf{r}}$, for which the local transformation between the two wave-function representations can be written as
\begin{equation}\label{transformation}
 \widetilde{\psi}_0\exp\left[i\mathbf{k}_{re}\cdot \mathbf{r}\right]\exp\left[\frac{1-n(\mathbf{r}_0)}{n(\mathbf{r}_0)}\right]\exp\left[\frac{-\delta \mathbf{r} \cdot \boldsymbol{\nabla} n}{n^2(\mathbf{r}_0)}\right] =\psi.
\end{equation}
Here $\mathbf{r}_0$ is the local reference point, and $\mathbf{k}_{re}$ denotes the real part of the wave vector associated with the Minkowski wave vector. The exponential attenuation factor indicates that the transformed wave function acquires an evanescent-wave-like character, i.e., it decays along the direction of $\delta \mathbf{r}$\cite{Bliokh2015TransverseAL}. Accordingly, the local wave vector becomes complex, $\mathbf{k}=\mathbf{k}_{re} + i\mathbf{k}_{im}$, with $\mathbf{k}_{im} = \nabla n /n^2$. From the viewpoint of the photon subsystem, the interaction with the inhomogeneous medium manifests itself as an effective imaginary contribution to the wave vector. Consequently, the local Abraham momentum operator takes the form
\begin{equation}\label{pabr}
    \hat{p}^{Abr} =\frac{\hbar\omega}{n} =\frac{\hat{p}^{Min}}{n^2} +i \frac{\nabla n}{n^3}.
\end{equation}
Here we have used $\hat{p}^{Min}=\hbar k_{re}$ and the corresponding expression for $k_{im}$.

From another perspective, in the $\psi$ representation with $\psi=\psi_0 e^{i\mathbf{k}'_{\mathrm{re}}\cdot\mathbf r}$, the local transformation between the two wave functions is given by
\begin{equation}
 \psi_0 \exp\left[i\mathbf{k}^{\prime}_{re}\cdot \mathbf{r}\right] \exp\left[n(r_0)-1\right]\exp\left[\delta \mathbf{r} \cdot \boldsymbol{\nabla} n\right] =\widetilde{\psi}.
\end{equation}
Here $\mathbf{k}^{\prime}_{re}$ is associated with the Abraham wave vector. In this case, the effective complex wave vector is $\mathbf{k}^{\prime}=\mathbf{k}^{\prime}_{re} - i\mathbf{k}^{\prime}_{im}$, with $\mathbf{k}^{\prime}_{im} = \nabla n$.
Since the wave-vector--frequency relation in this representation is $k^{\prime} =\hbar \omega/n$, one obtains the corresponding Minkowski momentum operator as
\begin{equation}\label{pmin}
    (\hat{p}^{\prime})^{Min} =n\hbar\omega = n^2 (\hat{p}^{\prime})^{Abr} - in\nabla n.
\end{equation}
Here we have substituted $(\hat{p}^{\prime})^{Abr} =\hbar k^{\prime}_{re}$ and the corresponding expression for $k^{\prime}_{im}$. After integrating Eqs.~(\ref{pabr}) and (\ref{pmin}) one can see that the two representations lead to equivalent physical interpretations at the level of the total momentum. However, this equivalence does not generally hold locally: the corresponding local momentum densities in the two representations are not identical.

\subsection{Gauge-Field Connection between the Minkowski and Abraham Momenta}
We now focus on the gauge field generated by the anisotropy of the dielectric medium. The tensor $\boldsymbol{\epsilon}^{-1}_{\perp}$ maps the free-space optical spin basis $\mathbf{e}_{\sigma}$ onto the local medium-adapted basis $\mathbf{e}_{\sigma}(\mathbf{k})$. This basis transformation modifies the orthogonality and completeness relations of the spin basis, while leaving the momentum operator unchanged \cite{yang}. Since this contribution is not part of the effective gauge field considered here, we focus instead on the coupling between the longitudinal field components and the anisotropic medium response, characterized by ${\bf q} $. This coupling induces a gauge transformation between the two wave-function representations. More generally, in optical systems formulated in terms of the electric and magnetic fields $\mathbf{E}$ and $\mathbf{H}$(denoted by $\psi$) ,this contribution from the dielectric tensor $\epsilon$ also appears as a gauge-field term \cite{Gauge,Liu}. In the present optical Dirac formulation\cite{Feng2022,yang2025inducedberryconnectionphotonic}, described in terms of $\mathbf{D}$ and $\mathbf{B}$ (denoted by $\widetilde{\psi}$), the corresponding effective gauge field arises from the inverse dielectric tensor. We consider an inverse dielectric tensor of the form
\begin{equation}\label{medium}
    \epsilon^{-1}=\left(\begin{array}{ccc}
        1 & 0 & Q_{+1}\\
        0 & 1 & Q_{-1}\\
         Q_{-1} & Q_{+1} & 1+2Q_{+1}Q_{-1}
    \end{array}\right),
    \end{equation}
and by choosing appropriate values of the parameters $Q_{\pm1} $ one can see that this effective medium corresponds to an equivalent helical waveguide\cite{yang2025inducedberryconnectionphotonic}. Taking into account the relations $\mathbf{D} -\mathbf{P} =\mathbf{E}$ and $ \mathbf{B} -\mathbf{M} =\mathbf{H}$, it follows
that the polarization $\mathbf{P}$ and magnetization $\mathbf{M}$ of
the medium are described by the dipole parameters $Q_{\pm1}$.

The Minkowski momentum density is
\begin{equation}\label{S^Min}
    \mathbf{S}^{Min}=\mathbf{D}\times \mathbf{B} =\mathbf{D}_{T}\times \mathbf{B}_{T} + (\mathbf{D}_{\parallel} \times \mathbf{B}_{T} +\mathbf{D}_{T}\times \mathbf{B}_{\parallel}),
\end{equation}
where the subscripts $\parallel$ and $T $ denote the longitudinal and transverse field components, respectively. The Abraham momentum density is defined as
\begin{equation}\label{S^Abr}
    \mathbf{S}^{Abr}=\mathbf{E} \times \mathbf{H}=\mathbf{E}_{T}\times \mathbf{H}_{T} + (\mathbf{E}_{\parallel} \times \mathbf{H}_{T} +\mathbf{E}_{T}\times \mathbf{H}_{\parallel}),
\end{equation}
which can be expressed in terms of $\mathbf{D}$ and $\mathbf{B}$. The first term in $\mathbf{S}^{Abr}$ is given by
\begin{equation}\label{K_z}
    \mathbf{E}_{T}\times \mathbf{H}_T = (\mathbf{D}_{T}  \times \mathbf{B}_{T} ) + (\mathbf{q}D_z\times \mathbf{B}_{T} + \mathbf{D}_{T}  \times\mathbf{q}B_z) .
\end{equation}
The correction term in the above equation then yields
{\small\begin{equation}\label{qcdotk}
\begin{aligned}
&\int d^3r\,
\left(
\mathbf{q}D_z\times\mathbf{B}_{T}
+
\mathbf{D}_{T}\times\mathbf{q}B_z
\right)
=
-\int d^3r\,
\frac{\mathbf{q}\cdot\hat{\mathbf{k}}}{k_z}
\left(
\mathbf{D}_T\times\mathbf{B}_T
\right)
\\
&\qquad =
-\int d^3k
\sum_{\sigma}
\left(
\mathbf{q}\cdot\hat{\mathbf{k}}
\right)
\left(
a^{\dagger}_{\sigma} a_{\sigma}
+
b^{\dagger}_{\sigma} b_{\sigma}
\right).
\end{aligned}
\end{equation}}
In the first equality, the longitudinal components $D_z$ and $B_z$ are expressed in terms of the transverse components using Eq.~(\ref{transverse}). In the second equality, we use the fact that $(\mathbf D_T\times \mathbf B_T)\propto k_z$ in Eq.~(\ref{Min-momentum}), which cancels the factor of $k_z$ and allows the result to be written in terms of creation and annihilation operators.

To obtain the transverse part of $\mathbf{S}^{\rm Abr}$, we first determine the explicit form of $\epsilon$ and then employ the transversality conditions for $\mathbf{D}$ and $\mathbf{B}$. The longitudinal field components can be expressed in terms of the transverse components as
\begin{equation}
    E_z  = \frac{-1}{K_z}(\hat{K}_- E_+ + \hat{K}_+E_-),\quad H_z  = \frac{-1}{K_z}(\hat{K}_- H_+ + \hat{K}_+H_-).
\end{equation}
Here $\hat{K}_i=-iD_i$ denotes the momentum-operator representation of the covariant derivative, defined by
\begin{equation}\label{covariant-derivative}
    \hat{K}_{\pm} =\hat{k}_{\pm} - \hat{k}_zQ_{\pm 1} +(\hat{k}_{\mp}Q^2_{\pm 1}-\hat{k}_{\pm}Q^2_1), \quad \hat{K}_z = \hat{k}_z - (\hat{\mathbf{k}}\cdot\mathbf{q}), 
\end{equation}
In the weak-anisotropy approximation, terms of order $ O(Q^2)$ are negligible and will be omitted in the following analysis. Hence, the transverse part of the Abraham momentum is
\begin{equation}
\begin{aligned}\label{Abr-momentum}
    \int d^3 r \,(\mathbf{E}_{\parallel} \times \mathbf{H}_{T} +\mathbf{E}_{T}\times \mathbf{H}_{\parallel})=\int d^3 r \,\frac{\hat{\mathbf{K}}}{K_z}(\mathbf{E}_T\times\mathbf{H}_T)_z\\
    = \int d^3 k \sum_s\,\boldsymbol{K}_{\perp}\, (a^{\dagger}_{\sigma} a_{\sigma} + b^{\dagger}_{\sigma} b_{\sigma}).
    \end{aligned}
\end{equation}
In the last step, we use the relation between the longitudinal parts of the Abraham and Minkowski momentum densities, Eqs.~(\ref{K_z}) and (\ref{qcdotk}), to eliminate the factor of $K_z$, and rewrite the result in terms of creation and annihilation operators. 

From Eqs.~(\ref{Min-momentum}), (\ref{K_z}), (\ref{qcdotk}), 
(\ref{covariant-derivative}), and (\ref{Abr-momentum}), it can be seen that, in the $\widetilde{\psi}$ representation, the Abraham momentum density corresponds to the covariant derivative, while the Minkowski momentum density corresponds to the ordinary derivative. These two derivatives are related by
\begin{equation}
    D_i  =\partial_i - i\widetilde{A}_i, 
\end{equation}
where the effective composite gauge field is defined through $\widetilde{A}_z = (\hat{\mathbf{k}}\cdot \mathbf{q}) $ and  $\widetilde{A}_{\pm} =\hat{k}_zQ_{\pm 1}$. The transverse gauge-field components are governed by the material parameters $Q_{\pm1}$, while the $z$-component of the wave vector plays the role of an effective coupling coefficient. To make the comparison with electronic systems below more direct, we absorb this coefficient into the definition of the gauge field. Accordingly, for the medium specified by Eq.~(\ref{medium}), the effective transformation between the two wave-function representations associated with the vector $\mathbf{q}$ is $\psi \propto \exp[-i\int \widetilde{A}_i dx^i]\widetilde{\psi}.$ Upon integration, the Abraham and Minkowski momenta are related by $\mathbf{P}^{Abr} = \mathbf{P}^{Min} -\widetilde{\mathbf{A}}(\epsilon^{-1}).$

The connection between the two momenta and the effective gauge field can also be formulated in the wave-function representation $\psi$. In this representation, one starts from the dielectric tensor $\epsilon$, whose inverse is given in Eq.~\eqref{medium}, and repeats the above analysis with the Minkowski momentum expressed in terms of $\mathbf E$ and $\mathbf H$. To leading order, the effective gauge field takes an analogous form, $A_z\approx\hat{\mathbf{k}}\cdot \mathbf{q}^{\prime}$ and $\mathbf{A}_{\perp} \approx\hat{k}_z\mathbf{q}^{\prime} =\hat{k}_z(Q^{\prime}_{+1},Q^{\prime}_{-1})^T$, as defined in Eq.~\eqref{covariant-derivative}, with $Q^{\prime}_{\pm1}=-Q_{\pm1}$. Thus, the parameters $Q^{\prime}_{\pm1}$ act as gauge-field components in the wave-function formulation of Maxwell’s equations expressed in terms of $\mathbf{E}$ and $\mathbf{H}$\cite{Gauge,Liu}. When $\mu \neq \epsilon$, this gauge field generally becomes non-Abelian \cite{Chen2018NonAbelianGF}.  In $\psi$ representation, the Minkowski momentum density is associated with the covariant derivative, whereas the Abraham momentum density is associated with the ordinary derivative. The two momentum operators are connected by an effective gauge field $\mathbf{A}(\epsilon)$, as summarized in Table~\ref{table1}. Although the corresponding local momentum densities are representation dependent, spatial integration leads to the same relation between the total Abraham and Minkowski momenta in the two representations. Accordingly, the two wave functions are related by a gauge transformation of the form $\widetilde{\psi} \propto \exp[-i\int A_i dx^i]\psi$.
\begin{table}[ht]
\centering
\footnotesize
\renewcommand{\arraystretch}{1.2}

\begin{tabular}{lcc}
\toprule
\textbf{Wavefunction}
&
\textbf{Min Momentum}
&
\textbf{Abr Momentum }
\\
\midrule

$\widetilde{\psi}(\mathbf D,\mathbf B)$
&
$P_i^{Min}\leftrightarrow \partial_i$
&
$P^{Abr}_i\leftrightarrow D_i=\partial_i-i\widetilde{A}_i$
\\

$\psi(\mathbf E,\mathbf H)$
&
$P_i^{Min}\leftrightarrow D_i=\partial_i-iA_i$
&
$P^{Abr}_i\leftrightarrow \partial_i$
\\

\bottomrule
\end{tabular}
\caption{Gauge-theoretic correspondence between Minkowski and Abraham momentum densities in the $\widetilde{\psi}$ and $\psi$ representations.
}
\label{table1}
\end{table} 

\section{Photon Dynamics and the Spin--Zeeman Coupling}
In quantum mechanics, the generator of ordinary spatial translations is the canonical momentum, which is represented by the ordinary derivative, whereas the kinetic momentum is represented by the covariant derivative. Therefore, under the quantum--classical correspondence, the Minkowski and Abraham momenta can be identified as the canonical and kinetic momenta of photons, respectively. To make this correspondence more explicit, we formulate the effective classical Hamiltonian for the photon system in the wave-function representation $\widetilde{\psi}$ of Maxwell's equations for the medium described by the dielectric tensor in Eq.~(\ref{medium}). The resulting optical Dirac-like evolution equation reads~\cite{Feng2022,yang2025inducedberryconnectionphotonic}
\begin{equation}\label{ham}
    i\frac{\partial \widetilde{\psi} }{\partial t} = (\gamma^0 H_0 +\gamma^{\perp} H_{\perp})\widetilde{\psi},
\end{equation}
where the $\gamma$ matrices are taken in the Dirac representation, with
\begin{equation}
\begin{aligned}
H_0
&= \hat{k}_z \boldsymbol{I}
+ \frac{1}{k_z}
\begin{pmatrix}
H^+_0 & 0 \\
0 & H^-_0
\end{pmatrix}, \\
H^+_0
&= \hat{k}_+ \hat{k}_-
- \hat{k}_z \left(\hat{k}_+ Q_{-1} + Q_{+1}\hat{k}_-\right), \\
H^-_0
&= \hat{k}_- \hat{k}_+
- \hat{k}_z \left(Q_{-1}\hat{k}_+ + \hat{k}_- Q_{+1}\right).
\end{aligned}
\end{equation}
and 
\begin{equation}
     H_{\pm}=\frac{1}{k_z}\bigl(\hat{k}^2_{\pm} -\hat{k}_z(Q_{\pm1}\hat{k}_{\pm} +  \hat{k}_{\pm}Q_{\pm1})\bigr).
\end{equation}
The off-diagonal terms couple the positive- and negative-energy sectors and therefore give rise to Zitterbewegung (ZB). The resulting Hamiltonian is structurally equivalent to the Feshbach--Villars Hamiltonian of a free scalar particle \cite{RevModPhys.30.24} under the correspondence $m \leftrightarrow k_z$, and consequently exhibits similar ZB dynamics for light
\cite{Unal1997,KOBE19997,10.1119/1.12819,PhysRevA.105.062211}. Moreover, it satisfies the pseudo-Hermiticity relation $\gamma^0 H \gamma^0 =H^{\dagger}$. Accordingly, the appropriate inner product is defined with respect to the metric operator $\gamma^0$. This pseudo-Hermitian structure naturally accounts for the additional $\gamma^0$ factor appearing in the expectation value of the momentum operator in Eq.~(\ref{Min-momentum}).

In the semiclassical picture, the photon kinetic momentum is defined by $\boldsymbol{\pi}_{\perp} = \mathbf{k}_{\perp} - \widetilde{\mathbf{A}}_{\perp}$ with $\widetilde{\mathbf{A}}_{\perp} =\mathbf{q}k_z$.
The Hamiltonian $H_0$ takes the form
\begin{equation}
    H_0= \hat{k}_z \mathbf{I} + \frac{(\boldsymbol{\sigma}_{\perp} \cdot \boldsymbol{\pi}_{\perp})^2}{k_z} -\hat{k}_zQ^2_{1},
\end{equation}
where $\boldsymbol{\sigma}_{\perp}$ represents the transverse components of the Pauli matrices. Using the Pauli algebra and the commutation relation
\begin{equation}\label{pi_ipi_j}
    [\pi_i,\pi_j] = i\epsilon_{ijk}B^{eff}_k,
\end{equation}
where the effective magnetic field is $\mathbf{B}^{eff} =\boldsymbol{ \nabla} \times \widetilde{\mathbf{A}}_{\perp}$, one finds $(\boldsymbol{\sigma}_{\perp} \cdot \boldsymbol{\pi}_{\perp})^2 = \boldsymbol{\pi}_{\perp}^2/2 - \sigma_z B_z^{eff}$, with $\boldsymbol{\pi}^2_{\perp}=(\pi^2_x+\pi^2_y)$. Therefore, the Hamiltonian $H_0$ can be rewritten as
\begin{equation}\label{spin-zeeman}
     H_0 =\hat{k}_z \mathbf{I} + \frac{1}{2k_z} \boldsymbol{\pi}_{\perp}^2  +V-\sigma_z B^z_{eff}/k_z ,
\end{equation}
with $V=-k_zQ^2_{1}$. The last term represents an effective spin-Zeeman
coupling that appears even without an external magnetic field. This structure closely parallels the Pauli Hamiltonian obtained as the nonrelativistic limit of the Dirac equation~\cite{PhysRev.78.29}.
\begin{figure*}[t]
    \centering
            \centering
            \includegraphics[height=0.5\textheight]{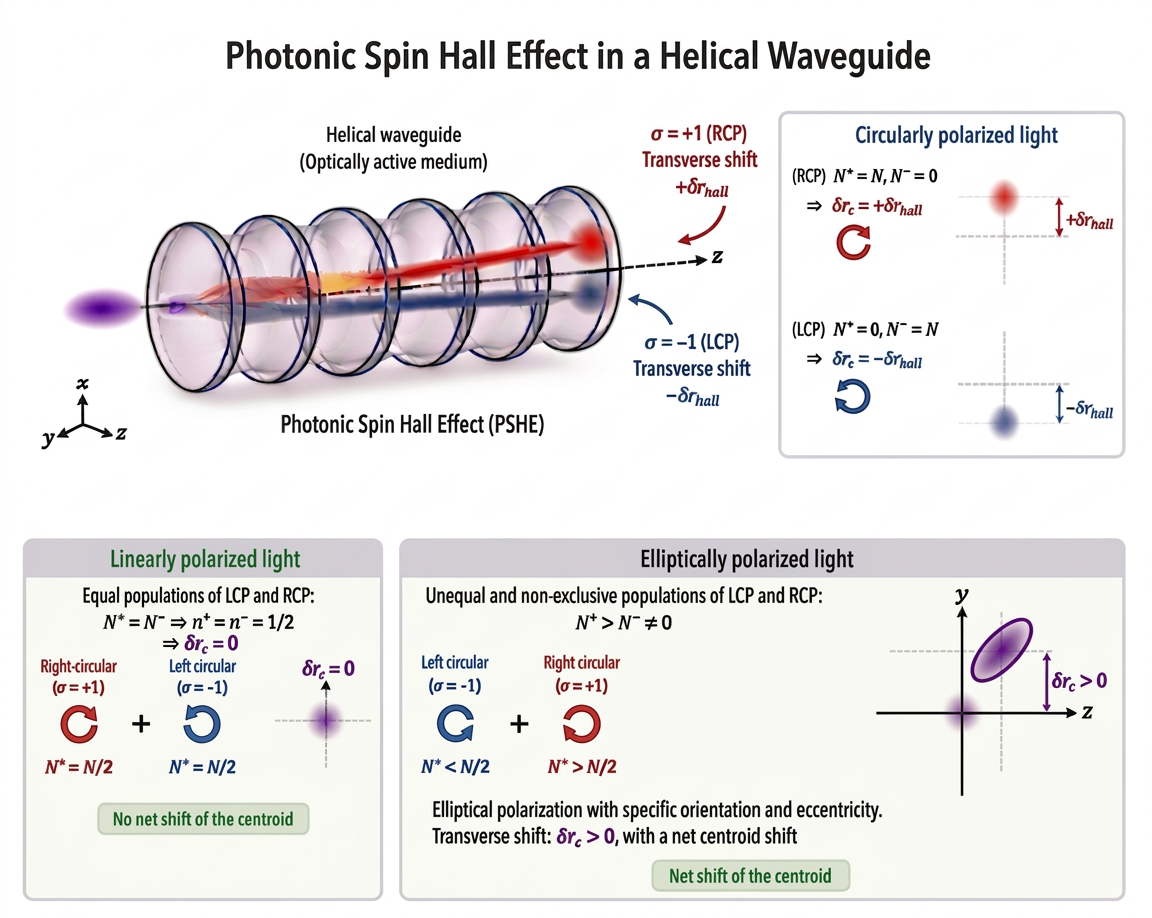}
             \caption{Schematic illustration of the optical spin Hall effect from the photon perspective. For linearly polarized light, the photon numbers of the two spin components are equal, so the centroid displacement vanishes, $\delta\mathbf{r}_c=0$.For elliptically polarized light, the imbalance between the two spin components leads to a partial displacement,$0<\delta\mathbf{r}_c<\mathbf{r}_{hall}$.For circularly polarized light, the displacement reaches its maximum,$\delta\mathbf{r}_c =\mathbf{r}_{hall}$. }
             \label{fig1}
\end{figure*}
In the effective helical waveguide \cite{yang2025inducedberryconnectionphotonic}, the covariant derivative describing the two-dimensional transverse dynamics takes the form
\begin{equation}
    D_i=\partial_i + i \sigma \widetilde{A}_i\quad  \widetilde{A}_i=-\frac{2\pi\tau r_i}{\lambda},
\end{equation}
Here, the effective magnetic field satisfies $B^{\rm eff}_z/k_z = \tau$, corresponding to the torsion of the waveguide. Within the paraxial approximation, one has $k_z \approx k = 2\pi/\lambda$, where $\lambda$ is the wavelength.

In this system, photons propagating through the helical optical fiber acquire a helicity-dependent geometric phase \cite{10.1098/rspa.1984.0023,Chiao1986ManifestationsOB,PhysRevLett.57.937}. Within the semiclassical framework, the associated Berry curvature generates an anomalous transverse velocity, leading to a helicity-dependent displacement and ultimately giving rise to the photonic spin Hall effect
\cite{HallPhysRevLett.93.083901,Bliokh2008GeometrodynamicsOS,Bliokh_2009,Bliokh2015,yang2025inducedberryconnectionphotonic}. Equivalently, this phenomenon can be understood from the spin--Zeeman term in Eq.~(\ref{spin-zeeman}), which embodies the spin--orbit interaction of light
\cite{PhysRevA.46.5199,BLIOKH2004181,Bliokh2008GeometrodynamicsOS,Bliokh_2009,Bliokh2015,yang2025inducedberryconnectionphotonic}. The resulting leading order dispersion relation, $\omega = k_z + \sigma \tau$ depends explicitly on the photon helicity, giving rise to helicity-dependent propagation constants and phase velocities $v_p^{(\sigma)} = \omega/k_z^{(\sigma)}$. In a photon picture, the accumulation of photons with different helicities manifests macroscopically as the photonic spin Hall effect. Let $N^+$ and $N^-$ denote the numbers of right- and left-handed photons in a beam, respectively. Assuming equal effective masses, the displacement of the photon center-of-mass coordinate $\mathbf{r}_c$ can be written as
\begin{equation}
    \delta \mathbf{r}_c =(n^+-n^-) \delta \mathbf{r}_{hall} , 
\end{equation}
where $ n^{\pm} =N^{\pm}/N$ are the photon number fractions, and $\delta \mathbf{r}_{hall}$ is the Hall displacement of a circularly polarized beam propagating in the helical waveguide\cite{Bliokh2008GeometrodynamicsOS,Bliokh_2009}. A schematic illustration is shown in Fig.~\ref{fig1}.
\begin{table}[ht]
\centering
\renewcommand{\arraystretch}{1.3}
\begin{tabular}{lcc}
\toprule
&
\textbf{Electronic system}
&
\textbf{Photonic system}
\\
\midrule

Coupling parameter
&
$e$
&
$\sigma$
\\

Canonical momentum
&
$\mathbf p^e$
&
$\mathbf P^{\rm Min}(\mathbf{k})$
\\

Kinetic momentum
&
$\boldsymbol{\pi}^e$
&
$\mathbf P^{\rm Abr}(\boldsymbol{\pi})$
\\

Gauge field
&
$\mathbf A^{\rm el}$
&
$\widetilde{\mathbf A}$
\\

Momentum relation
&
$\boldsymbol{\pi}^e
=
\mathbf p^e-e\mathbf A^{\rm el}$
&
$\mathbf P^{\rm Abr}
=
\mathbf P^{\rm Min}
-\sigma\widetilde{\mathbf A}$
\\

Gauge phase
&
$\exp\!\left[
\frac{i e}{\hbar}
\int A_i^{\rm el}dx^i
\right]$
&
$\exp\!\left[
\frac{i \sigma}{\hbar}
\int \widetilde A_i dx^i
\right]$
\\

\bottomrule
\end{tabular}
\caption{
Dual description of electronic and photonic systems.
}
\label{table2}
\end{table}
After eliminating the explicit time dependence, the system reduces to an effective two-dimensional model propagating along the z direction with and governed by a Schrödinger-type equation \cite{PhysRevLett.80.1888,Feng2022,yang2025inducedberryconnectionphotonic}. within the semiclassical approximation and neglecting Zitterbewegung, the effective dynamics of photons propagating along the $z$ axis can be described by the effective Hamiltonian
\begin{equation}
    H^{\prime}_0 = \frac{\boldsymbol{\pi}_{\perp}^2}{2m} + V,
\end{equation}
where the effective mass $m$ is determined by the longitudinal wave number $k_z$. This two-dimensional system is structurally analogous to the electronic system in an external magnetic field, as summarized in Table~\ref{table2}.

\section{Conclusion}

We show that the distinction between Minkowski and Abraham momenta reflects different quantization schemes of the photon–medium system. In a field-theoretic representation of the wave function $\widetilde{\psi}$, Minkowski momentum corresponds to the ordinary derivative, whereas Abraham momentum assumes a covariant-derivative form, linked via a medium-induced gauge field. Within this framework, the optical Dirac-like equation reveals a spin–Zeeman term that generates helicity-dependent phase velocities, providing a microscopic origin of the photonic spin Hall effect. In the classical limit, the effective two-dimensional photon Hamiltonian is formally dual to that of an electron in an external gauge potential, with Minkowski and Abraham momenta naturally corresponding to canonical and kinetic momenta, respectively. 

\section*{Acknowledgements}
This work was supported by National Key R\&D Program
of China (grant No. 2024YFE0109802), the National Natural
Science Foundation of China (grants Nos. 12175320,
12375084), the Natural Science Foundation of Guangdong
Province, China (Grant No. 2022A1515010280).


\bibliographystyle{elsarticle-num} 
\bibliography{main}






\end{document}